# An X-ray Outburst from the Rapidly Accreting Young Star That Illuminates McNeil's Nebula

J.H. Kastner[A], M. Richmond[A], N. Grosso[B], D.A. Weintraub[C], T. Simon[D], A. Frank[E], K. Hamaguchi[F], H. Ozawa[B], A. Henden[G]

*A) Rochester Institute of Technology, Rochester, NY, USA; B) Laboratoire d'Astrophysique de Grenoble, Université Joseph-Fourier, France; C) Vanderbilt University, Nashville, TN, USA; D) Institute for Astronomy, Honolulu, HI , USA; E) University of Rochester, Rochester, NY, USA; F) Goddard Space Flight Center, Greenbelt, MD, USA; G) U.S. Naval Observatory, Flagstaff, AZ, USA*

**Young, low-mass stars are luminous X-ray sources[1] whose powerful X-ray flares[2-6] may exert a profound influence over the process of planet formation[7]. The origin of such emission is uncertain. Although many or perhaps most recently formed, low-mass stars emit X-rays as a consequence of solar-like coronal activity[1,8,9], it has also been suggested that X-ray emission may be a direct result of mass accretion onto the forming star[10-12]. Here we report X-ray imaging spectroscopy observations which reveal a factor ~50 increase in the X-ray flux from a young star that is presently undergoing a spectacular optical/IR outburst[13,14]. The outburst is thought to be due to the sudden onset of a phase of rapid accretion[14-16]. The coincidence of a surge in X-ray brightness with the optical/IR eruption demonstrates that strongly enhanced high-energy emission from young stars can occur as a consequence of high accretion rates. We suggest that such accretion-enhanced X-ray emission from erupting young stars may be short-lived, because intense star-disk magnetospheric interactions are quenched rapidly by the subsequent accretion flood.**

In November 2003 a young, low-mass star deeply embedded in the Lynds 1630 dark cloud, which is located in the M78 (NGC 2068) star formation region in Orion, brightened suddenly [13, 14] in a manner that is reminiscent of optical outbursts associated with certain, less cloud-obscured, pre-main sequence (pre-MS) stars [16-18]. Such rarely-observed eruptions indicate that, for most (if not all) low-mass, pre-MS stars, long periods of relatively quiescent mass accretion at rates of $10^{-7}$ solar masses $yr^{-1}$ or less are punctuated by intense, short-lived (<100 yr duration) periods during which such stars may accumulate as much as 0.01 solar masses [16].

It is possible that pre-MS accretion processes also generate X-rays. Such high-energy radiation would occur due to the mediation of the accretion disk and stellar magnetospheres or, perhaps, due directly to the disk itself [10]. Models predict either episodic or steady reconnection of magnetic field lines occurring at the strong shear layer wherein the stellar magnetosphere truncates the inner part of the accretion disk. As the fields are wound up due to the shear they can balloon above and below the disk, eventually leading to reconnection and flares. This star-disk reconnection mechanism was proposed to explain the repeating X-ray flares of the low-mass protostar YLW 15 [19]. Such activity may also drive well-collimated outflows or jets [20, 21].

As the first pre-MS star to be observed intensively just at the onset of an optical/IR outburst (Fig. 1), the erupting object in L1630 affords a unique opportunity to trace the X-ray luminosity and spectral evolution of what is, most likely, a major pre-MS accretion event. We serendipitously detected the erupting pre-MS star in X-rays in Nov. 2002 -- approximately 1 year before outburst -- during a deep observation of the L1630 region [22] with the Advanced CCD Imaging Spectrometer (ACIS) aboard the Chandra X-ray Observatory (CXO). In March 2004, following the optical outburst, we twice used CXO/ACIS to obtain short-exposure observations of a field centered on the erupting object, and detected a large increase in flux from the spatially coincident X-ray source (Table 1).

In all three CXO/ACIS observations, a point-like X-ray source is detected within 0.2" of the position of the erupting young star, which lies at the apex of a fan-shaped reflection nebulosity (McNeil's Nebula; Fig. 2). The sharp increase in X-ray flux post-outburst relative to pre-outburst closely tracks that of the optical and infrared brightness increase of the point-like optical/IR source (Fig. 1). Furthermore, the change in X-ray flux evidently was accompanied by a change in X-ray spectral hardness, as reflected in the mean energies of events at the three epochs of observation (Table 1). The results suggest that the source X-ray spectrum hardened considerably near the peak of the optical outburst, then softened as the outburst proceeded. The observed relationship between spectral hardness and X-ray flux indicates that the initial rise and subsequent decline in X-ray count rate is probably not due to changes in absorbing column along our line of sight but, instead, reveals large variations in the intrinsic source X-ray luminosity and temperature. Similarly, near-infrared monitoring suggests that most (~2.5 magnitudes) of the abrupt ~3 magnitude change in brightness at these wavelengths can be attributed to the object's increase in intrinsic luminosity, rather than to a large change in extinction along our line of sight [15].

Modeling of the X-ray spectrum obtained 2004 March 7 indicates that the temperature of the X-ray-emitting gas ($T_X$) was ~ $6 \times 10^7$ K and that the source is viewed through an absorbing column ($N_H$) of ~ $6 \times 10^{22}$ cm$^{-2}$ (Fig. 3). Given a gas-to-dust ratio typical of the interstellar medium, the best-fit value of $N_H$ corresponds to dust extinction of $A_V$ ~ 35 mag, whereas near-infrared measurements [15] yield $A_V$ ~ 15 mag. This discrepancy suggests that the gas-to-dust ratio in the circumstellar environment of the young star is larger than interstellar.

Assuming no change in the absorbing column toward the X-ray source, analysis of the pre- and post-outburst X-ray observations indicates that the intrinsic source X-ray luminosity rose from $L_X \sim 3x10^{29}$ erg/s pre-outburst to $\sim 10^{31}$ erg/s post-outburst (where we also assume the distance to L1630 is 400 pc [23] and that the pre-outburst temperature of the source was $T_X = 10^7$ K). This factor of ~30 increase is very similar to the increase in near-infrared luminosity [15], and its timing furthermore indicates a direct connection to the optical/IR event. In support of this conclusion, we note that about 15% of the ~1000 X-ray-emitting young stars in the Orion Nebula Cluster (ONC) exhibited X-ray flaring (in two CXO observations spanning a combined 83 ks period [24]); of these, only about five exhibited count rate increases of a factor 10 or more. Thus, based solely on the most highly variable ONC X-ray sources, the probability that we have observed a large-amplitude X-ray flare that is unrelated to the optical/IR outburst is < 3%. Furthermore, the erupting pre-MS star in L1630 displayed no evidence of strong X-ray variability during the long, pre-outburst observation in November 2002.

The pre-outburst spectral energy distribution, circumstellar dust mass, and bolometric luminosity of the object in L1630 resemble those of T Tauri stars and erupting pre-MS (FU Ori) stars [25], and its outburst light curve to date (Fig. 1) is very similar to those of the FU Ori prototypes at outburst onset [16, 17]. Hence, it seems reasonable to attribute the rapid optical/IR brightness increase of this source to a sharp transition from relatively slow to exceedingly rapid accretion [14-16]. The contemporaneous X-ray burst therefore is best explained as ultimately due to accretion, as well. The inferred post-outburst X-ray temperature was far too high for the X-rays to be generated by shocks resulting from accretion onto a low-mass, pre-MS star, however [12]. Instead, the burst of X-rays was most likely generated via star-disk magnetic reconnection events that occurred in conjunction with such mass infall. This process may also launch new, collimated outflows or jets. Indeed, prior to its recent eruption, the pre-MS star in L1630 had been identified as the exciting source of a chain of extended emission nebulosity that appears to terminate at a shock-excited Herbig-Haro object (HH 23) [26]. The presence of these structures suggests that the present optical/IR/X-ray outburst from this object may be merely the latest of a series of such events.

If FU Ori stars are indeed among the most rapidly accreting pre-MS stars [16], and X-rays from pre-MS stars can be ascribed in part to accretion, then one would naively expect all FU Ori stars to be luminous pre-MS X-ray sources. It is noteworthy, then, that prior to the observations reported here only two FU Ori candidates, Z CMa and L1551 IRS5, had been detected in X-rays; furthermore, both exhibit very low $L_X/L_{bol}$ ratios [27, 28] of ~ $10^{-6}$, compared with $L_X/L_{bol}$ ~ $10^{-3}$ for the erupting young star in L1630 (as estimated near the peak of its outburst). This suggests that the very large accretion rates during the steady-state phase following an FU Ori outburst eventually push the star-disk boundary sufficiently close to the stellar photosphere that the accretion becomes non-magnetospheric [29], thereby effectively "quenching" X-ray emission from such objects long before the rapid accretion phase itself subsides. The precipitous drop in the X-ray flux and spectral hardness of the erupting L1630 pre-MS star, post-outburst, may signal the onset of this quenching phase, or it may indicate that the abrupt change in the nature of the star-disk interactions has triggered a phase of strong variability in both X-ray luminosity and temperature.

[1] E.D. Feigelson, T. Montmerle. High-Energy Processes in Young Stellar Objects, Ann. Rev. Astron. Astrophys., 37, 363-408 (1999)

[2] T. Preibisch, R. Neuhaeuser, J.M. Alcala. A giant X-ray flare on the young star P1724, Astron. Astrophys. 304, L13-L16 (1995)

[3] K. Koyama, K. Hamaguchi, S. Ueno, N. Kobayashi, E.D. Feigelson. Discovery of Hard X-Rays from a Cluster of Protostars, Pub. Astron. Soc. Japan, 48, L87-L92 (1996)

[4] N. Grosso, T. Montmerle, E.D. Feigelson, P. Andre, S. Casanova, J. Gregorio-Hetem. An X-ray superflare from an infrared protostar, Nature, 387, 56-58 (1997)

[5] H. Ozawa, F. Nagase, Y. Ueda, T. Dotani, M. Ishida. Detection of Hard X-Rays from a Class I Protostar in the HH 24-26 Region in the Orion Molecular Cloud, Astrophys. J., 523, L81-L84 (1999)

[6] K. Imanishi, K. Koyama, Y. Tsuboi. Chandra Observation of the rho Ophiuchi Cloud, Astrophys. J., 557, 747-760 (2001)

[7] E.D. Feigelson, G.P. Garmire, S. Pravdo. Magnetic Flaring in the Pre-Main-Sequence Sun and Implications for the Early Solar System, Astrophys. J., 572, 335-349 (2002)

[8] J.H. Kastner, D.P. Huenemoerder, N.S. Schulz, C.R. Canizares, J. Li, D.A. Weintraub. The Coronal X-Ray Spectrum of the Multiple Weak-lined T Tauri Star System HD 98800, Astrophys. J., 605, L49-L52 (2004)

[9] K. G. Stassun, D. R. Ardila, M. Barsony, G. Basri, R. D. Mathieu. X-ray Properties of Pre-Main-Sequence Stars in the Orion Nebula Cluster with Known Rotation Periods, Astron. J., in press; astro-ph/0403159 (2004)

[10] F.H. Shu, H. Shang, A.E. Glassgold, T. Lee. X-rays and fluctuating X-winds from protostars, Science, 277, 1475-1479 (1997)

[11] J.H. Kastner, D.P. Huenemoerder, N.S. Schulz, C.R. Canizares, D.A. Weintraub. Evidence for Accretion: High-Resolution X-Ray Spectroscopy of the Classical T Tauri Star TW Hydrae, Astrophys. J., 567, 434-440 (2002)

[12] B. Stelzer, J. H. M. M. Schmitt. X-ray emission from a metal depleted accretion shock onto the classical T Tauri star TW Hya, Astron. Astrophys., in press; astro-ph/0402108 (2004)

[13] J. W. McNeil, IAU Circular 8284, 2004 February 9 (2004)

[14] C. Briceno, A.K. Vivas, J. Hernandez, N. Calvet, L. Hartmann, T. Megeath, P. Berlind, M. Calkins, S. Hoyer. McNeil's Nebula in Orion: The Outburst History, Astrophys. J. 606, L123-L126 (2004)

[15] B. Reipurth, C. Aspin. IRAS 05436-0007 and the Emergence of McNeil's Nebula, Astrophys. J., 606, L119-L122 (2004)

[16] L. Hartmann & S.J. Kenyon. The FU Orionis Phenomenon, Ann. Rev. Astron. Astrophys., 34, 207-240 (1996)

[17] G.H. Herbig. Eruptive phenomena in early stellar evolution, Astrophys. J., 217, 693-715 (1977)

[18] G.H. Herbig, C. Aspin, A.C. Gilmore, C.L. Imhoff, A.F. Jones. The 1993-1994 Activity of EX Lupi, Pub. Astron. Soc. Pac., 113, 1547-1553 (2001)

[19] T. Montmerle, N. Grosso, Y. Tsuboi, K. Koyama. Rotation and X-Ray Emission from Protostars, Astrophys. J., 532, 1097-1110 (2000)

[20] M. Hayashi, K. Shibata, R. Matsumoto. X-Ray Flares and Mass Outflows Driven by Magnetic Interaction between a Protostar and Its Surrounding Disk, Astrophys. J., 468, L37-L40 (1996)

[21] A.P. Goodson, R.M.Winglee, K.-H. Boehm. Time-dependent Accretion by Magnetic Young Stellar Objects as a Launching Mechanism for Stellar Jets, Astrophys. J., 489, 199-209 (1997)

[22] T. Simon, S. M. Andrews, J. T. Rayner, S. A. Drake. X-Ray and Infrared Observations of Embedded Young Stars in L1630, Astrophys. J., in press; astro-ph/0404260 (2004)

[23] Anthony-Twarog, B.J. The H-beta distance scale for B stars - The Orion association, Astron. J., 87, 1213-1222 (1982)

[24] E.D. Feigelson, P. Broos, J.A. Gaffney III, G. Garmire, L.A. Hillenbrand, S.H. Pravdo, L. Townsley, Y. Tsuboi. X-Ray-emitting Young Stars in the Orion Nebula, Astrophys. J., 574, 258-292 (2002)

[25] P. Ábrahám, Á. Kóspál, Sz. Csizmadia, A. Moór, M. Kun, G. Stringfellow. The infrared properties of the new outburst star IRAS 05436-0007 in quiescent phase, Astron. Astrophys. (in press), astro-ph/0404249 (2004)

[26] J. Eisloffel, R. Mundt. Parsec-Scale Jets From Young Stars, Astron. J., 114, 280-287 (1997)

[27] H. Zinnecker, Th. Preibisch. X-ray emission from Herbig Ae/Be stars: A ROSAT survey, Astron. Astrophys., 292, 152-164 (1994)

[28] F. Favata, C. V. M. Fridlund, G. Micela, S. Sciortino, A.A. Kaas. Discovery of X-ray emission from the protostellar jet L1551 IRS5 (HH 154) Astron. Astrophys., 386, 204-210 (2002)

[29] L. Hartmann. *Accretion Processes in Star Formation*, Cambridge University Press (1998)

[30] J. S. Kaastra, R. Mewe, H. Nieuwenhuijzen. Spex: a New Code for Spectral Analysis of X and UV Spectra, in *UV and X-ray Spectroscopy of Astrophysical and Laboratory Plasmas*, ed. K. Yamashita & T. Watanabe (Tokyo : Univ. Acad. Press), p. 411 (1996)

Correspondence and requests for materials should be addressed to J.H.K. (e-mail: jhk@cis.rit.edu).

with ESO's VLT at the Paranal Observatories under program ID 272.C-5045. The authors thank Bruce L. Gary for communicating results of I-band monitoring of the source.

Fig. 1: Near-infrared and X-ray photometry of the erupting young star in L1630 obtained in the period from late 1998 through 2004 March. CXO points are observed source fluxes calculated from the count rates in Table 1. The J- (1.25 μm), H- (1.65 μm), and K-(2.2 μm) band photometric data are from the 2MASS archive (pre-outburst) and SAAO SIRIUS, Gemini NIRI, and NOFS 1.55m (post-outburst); I-band photometric data are from Briceno et al. (2004 [14]) and B. Gary (2004, private comm.)

Fig. 2: Chandra X-ray Observatory and visible-light images of the region surrounding McNeil's Nebula. The Chandra/ACIS image (color) is overlaid with contours from a short-exposure (10 s) R-band image of the nebula obtained with the FORS2 camera at the European Southern Observatory's Very Large Telescope. The VLT/FORS2 and CXO/ACIS images were obtained on 2004 Feb. 18 and March 7, respectively. The Chandra image has red, green, and blue color coding for photons in the 0.5-1.5, 1.5-2.4, and 2.4-8.0 keV energy bands, respectively. Note the predominance of hard band X-rays from the (blue) source 3 = CXO J054613.1-000604 (J2000 coordinates 05h46m13.14s, -00d06m04.6s), which lies at the apex of McNeil's Nebula. This source is spatially coincident with the IR sources 2MASS J05461313-0006048 and IRAS 05436-0007 (= LMZ 12), which have been identified with the erupting young star now illuminating the nebula [14, 15]. Names (and J2000 coordinates) for the other sources in the field are as follows: source 1 = CXO J054619.4-000519 (05h46m19.47s, -00d05m19.7s; 2MASS J05461946-0005199, LkHα301); source 2 = CXO J054618.8-000537 (05h46m18.89s, -00d05m37.9s; 2MASS J05461889-0005381); and source 4 = CXO J054611.6-000627 (05h46m11.61s, -00d06m27.8s; 2MASS J05461162-0006279).

Fig. 3: Chandra X-ray Observatory spectrum of the source associated with McNeil's Nebula. Top: CXO/ACIS spectrum of CXO J054613.1-000604 obtained 2004 March 7 (points; errors are 1-sigma), with best-fit absorbed, optically thin thermal plasma model [30] overlaid (histogram). The best-fit model has a temperature $T_X$ = 5.6x10$^7$ K (with a lower limit of 1.5x10$^7$ K at 90% confidence, and unbounded upper limit) and intervening absorbing column $N_H$ = 5.7x10$^{22}$ cm$^{-2}$ (with a 90% confidence interval of (3.0-10)x10$^{22}$ cm$^{-2}$). Bottom: the residuals of the fit, in units of the uncertainties in the individual data points. This spectral analysis was performed with XSPEC version 11.3.

Table 1. **X-ray properties of the erupting young star and nearby, field sources.**

| # | November 14, 2002 | | March 7, 2004 | | March 22, 2004 | |
|---|---|---|---|---|---|---|
| | Count rate[a] [cts ks-1] | Median E[b] [keV] | Count rate[a] [cts ks-1] | Median E[b] [keV] | Count rate[a] [cts ks-1] | Median E[b] [keV] |
| 1 | 8.1 ± 0.4 | 1.4 ± 0.1 | 8.7 ± 1.4 | 1.6 ± 0.2 | 5.7 ± 1.0 | 1.9 ± 0.3 |
| 2 | 32.0 ± 0.8 | 1.4 ± 0.1 | 26.2 ± 2.3 | 1.4 ± 0.1 | 25.1± 2.2 | 1.5 ± 0.1 |
| 3[c] | 0.20 ± 0.06 | 2.5 ± 0.5 | 10.7 ± 1.4 | 3.6 ± 0.2 | 2.2 ± 0.6 | 2.0 ± 0.3 |
| 4 | 1.9 ± 0.2 | 1.7 ± 0.1 | 5.8 ± 1.1 | 1.7 ± 0.2 | 1.4 ± 0.5 | 1.1 ± 0.4 |

a) Count rates and statistical errors within 1.5" radius aperture. Nov. 14, 2002 data obtained in 55.9 ks effective exposure time with front-illuminated CCD ACIS-S2 [23]; March 7 and 22, 2004 data obtained in 5.5 ks and 4.9 ks exposure times, respectively, with back-illuminated CCD ACIS-S3. ACIS has a pixel size of 0.49 arcsec and the Chandra/ACIS combination is sensitive over the energy range 0.3-10 keV. The data were subject to standard processing by Chandra X-ray Center (CXC) pipeline software (version 7.1.1), and data reduction and analysis was performed using the CXC's CIAO package (version 3.0.2 and CALDB 2.26).

b) Median energy of source photons within energy range 0.5 keV to 8 keV, and statistical error on the median.

c) Source associated with McNeil's Nebula.

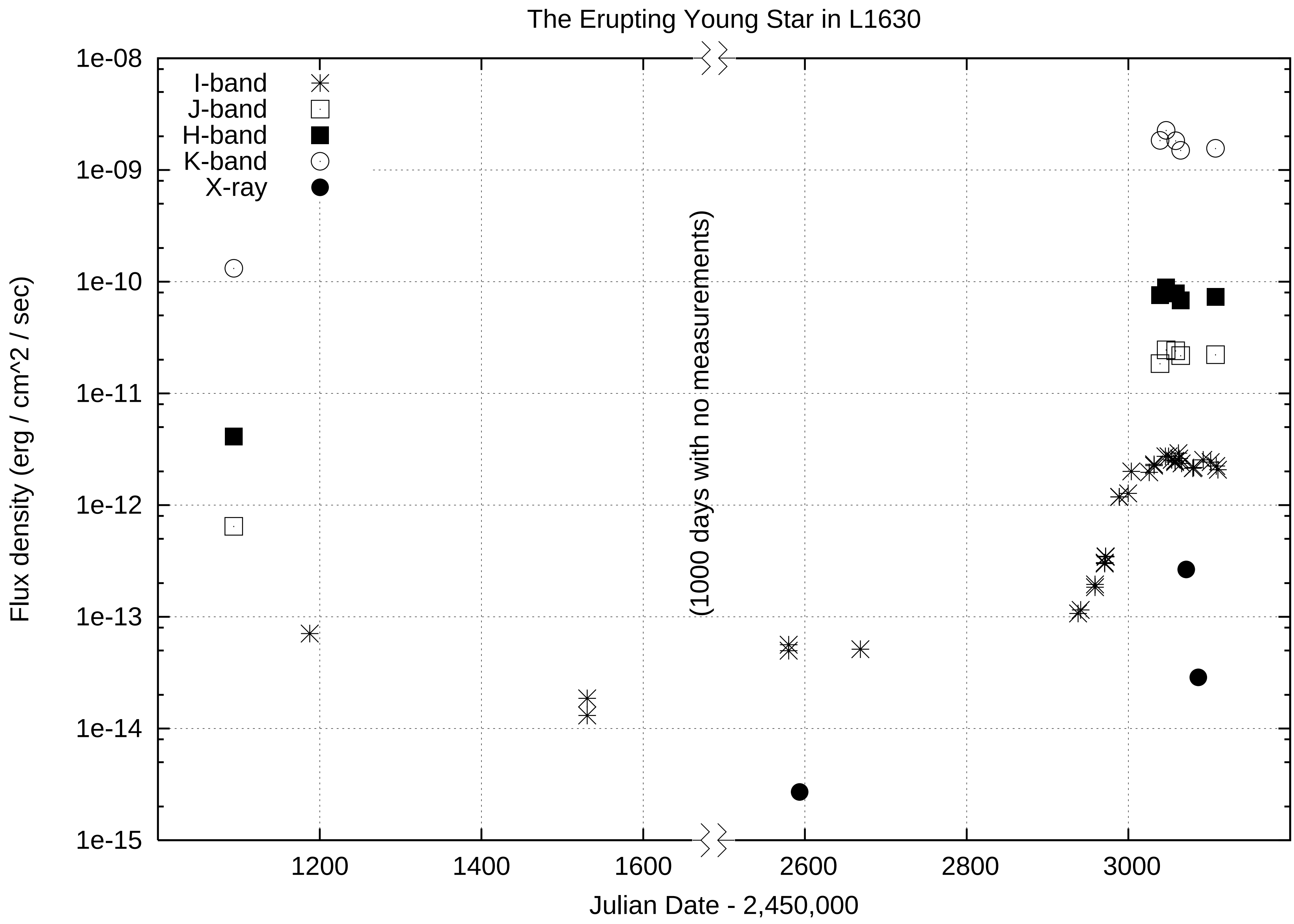

The Erupting Young Star in L1630
I-band
J-band
H-band
K-band
X-ray
(1000 days with no measurements)
Flux density (erg / cm^2 / sec)
1e-08
1e-09
1e-10
1e-11
1e-12
1e-13
1e-14
1e-15
1200
1400
1600
2600
2800
3000
Julian Date - 2,450,000

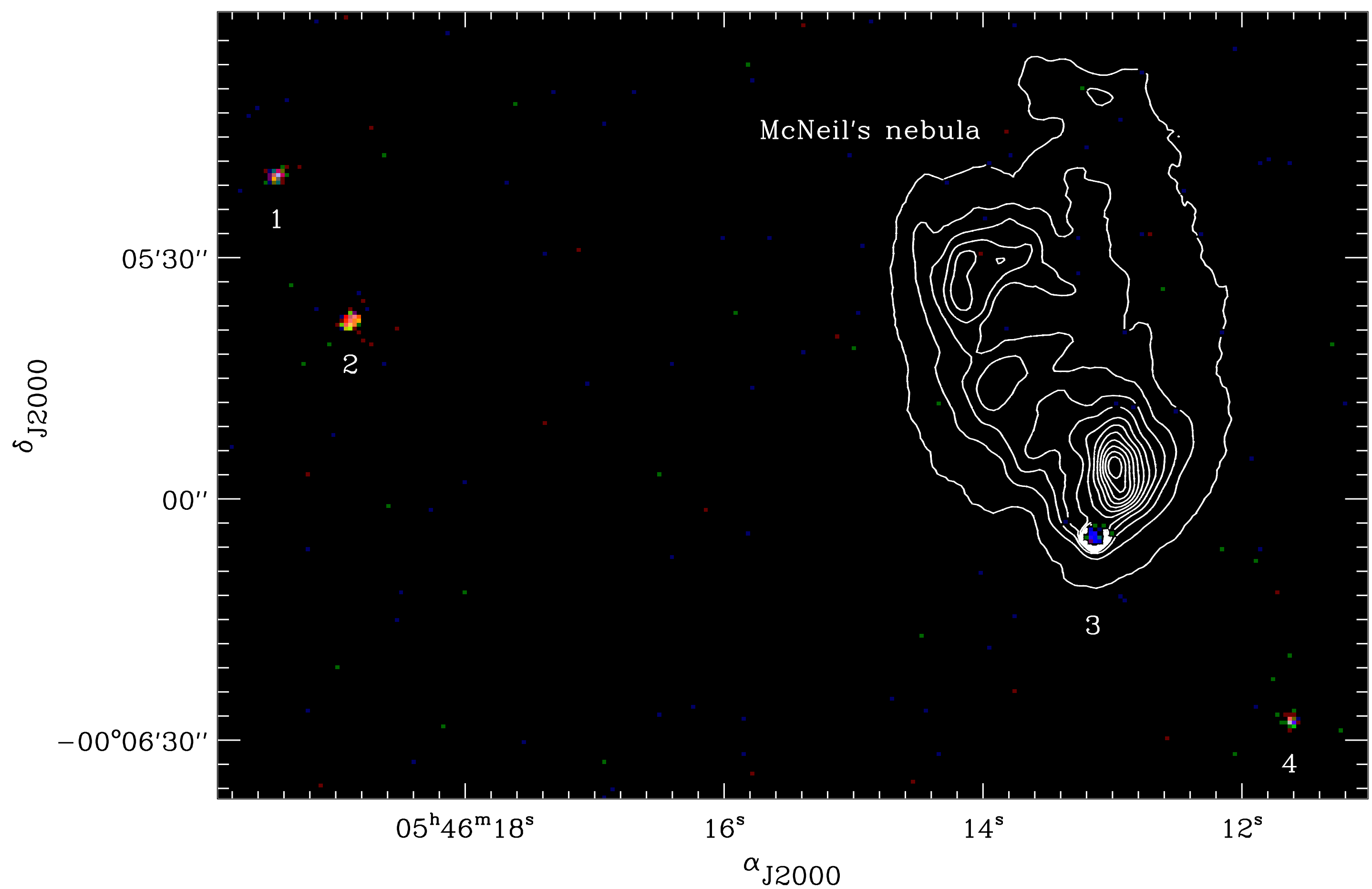
McNeil's nebula
1
2
3
4
05′30″
00″
−00°06′30″
$\delta_{J2000}$
05h46m18s
16s
14s
12s
$\alpha_{J2000}$

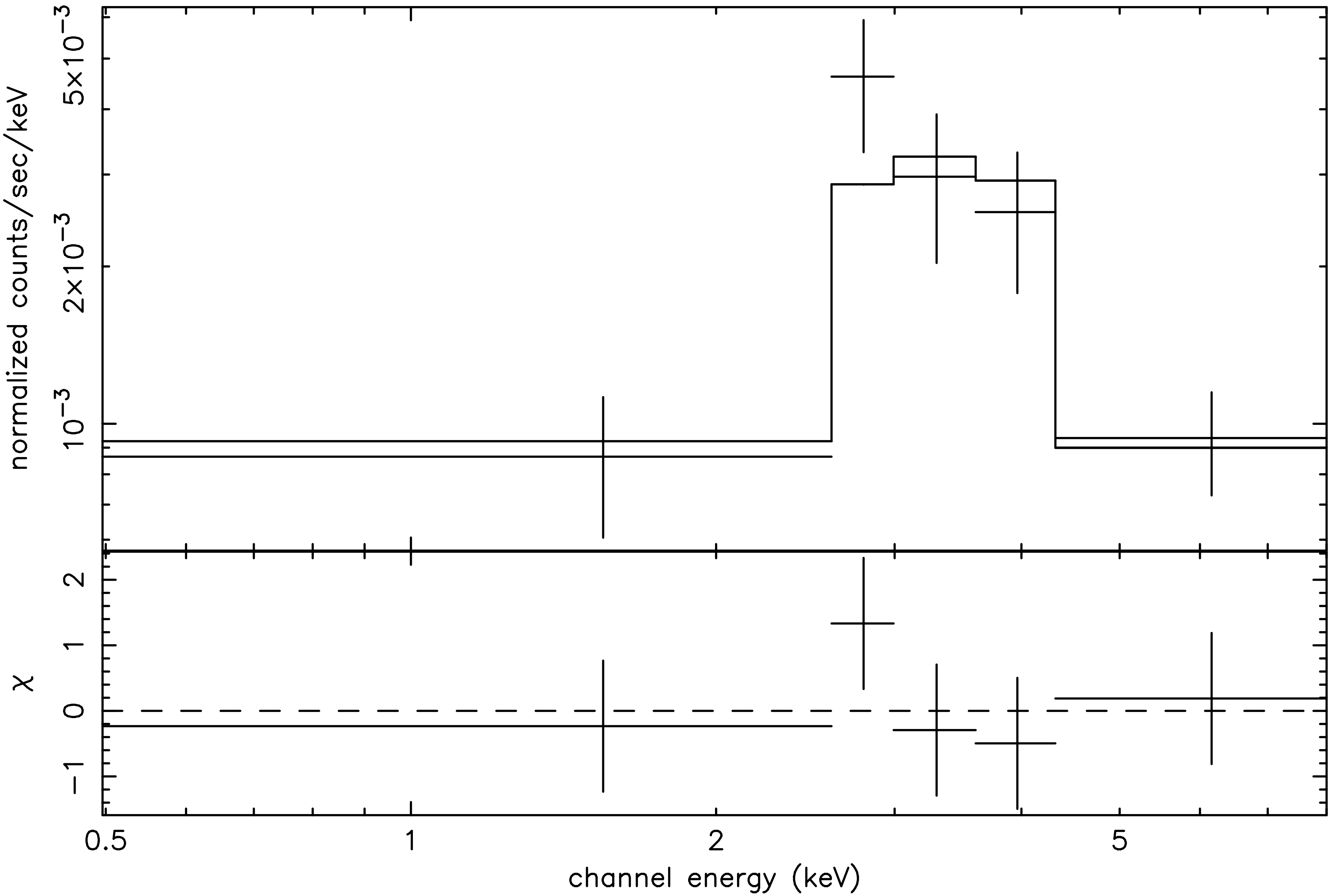
normalized counts/sec/keV
5×10⁻³
2×10⁻³
10⁻³
χ
2
1
0
−1
0.5
1
2
5
channel energy (keV)